\documentstyle[aps,prl,epsf,twocolumn]{revtex}
\begin{document}
\draft
\twocolumn[\hsize\textwidth\columnwidth\hsize\csname @twocolumnfalse\endcsname
\title{Extended state floating up in a lattice model:
Bona fide levitation fingerprints, irrespective of the correlation length}
 
\author{Ana L. C. Pereira and P. A. Schulz }
 
\address{Instituto de F\'{\i}sica Gleb Wataghin, Universidade Estadual 
de Campinas,  13083-970\\
Cx.P. 6165 Campinas, S\~ ao Paulo, Brazil}
 
\maketitle
 
\date{today}
 
\begin{abstract}
The evolution of extended states with magnetic field and disorder intensities
is investigated for 2D lattice models. The floating-up picture is revealed 
when the shift of the extended state, relative to the density of states, is 
properly taken into account, either for white-noise or correlated disorder.

\noindent PACS number(s)  73.43.Nq, 72.15.Rn, 71.30.+h
\end{abstract}

%\newpage
%\twocolumn
\vspace*{0.5cm}
]
\narrowtext
\section{Introduction}

The behavior of extended (current carrying) states in two dimensional 
(2D) disordered systems is the key feature in the transition between 
two widely accepted limits: absence of delocalized states without magnetic 
field and the existence of them in the Quantum Hall regime. Laughlin [1] 
and Khmelnitzkii [2] proposed that the extended states at the center of 
the Landau bands should float up in energy above the Fermi level. 
Although appealing, this levitation,  as the transition mechanism 
between the mentioned limits, is still controversial [3]. Perturbative 
approaches identify a weak levitation regime in the strong magnetic 
field limit [4, 5], while several numerical works report contradictory 
results. We mention some illustrative  controversial works based on 
tight-binding lattice models. The non-float-up picture, where the extended 
states are supposed to disappear at finite $B$ or disorder strength [6, 7], 
was recently reaffirmed in a proposed non-float-up phase diagram [8]. Other 
studies show evidences of floating up for white-noise disorder 
[9, 10, 11] with further disappearance of current carrying states, due 
to merging with states carrying negative Hall conductance (states of 
opposite Chern numbers: a lattice effect). A few very recent works have 
considered models with correlated disorders [12, 13] concluding that 
correlations would allow the floating up process not observed in 
white-noise case. Therefore, we revisit the problem, within the same 
framework of a 2D tight-binding lattice. The difference is that  
a direct comparison between different disorder models is established. 
We pay special attention to the deviation of the extended states 
relative to density of states (DOS) peaks in order to characterize 
the floating-up (as already suggested in references [9, 11]) and it was 
possible to observe and quantify bona fide levitation both in energy and 
in filling factor.

\section{Model calculation} 

The model Hamiltonian describes a square lattice of {\it s}-like orbitals, 
with nearest-neighbor interactions only: 
\begin{equation}
H = \sum_{i} \varepsilon_{i} c_{i}^{\dagger} c_{i} + \sum_{<i,j>} V 
(e^{i\phi_{ij}} c_{i}^{\dagger} c_{j} + e^{-i\phi_{ij}} c_{j}^{\dagger} c_{i}) 
\end{equation}

\hspace{-\parindent}where $c_{i}$ is the fermionic operator on site $i$.
The magnetic field is introduced by means of the phase 
$\phi_{ij}= 2\pi(e/h) \int_{j}^{i} \mathbf{A} \! \cdot \! d \mathbf{l} \;$ in 
the hopping parameter $V$. Considering the Landau gauge, $\phi_{ij}\!=\!0$ 
along the $x$ direction and $\phi_{ij}\!=\pm 2\pi (x/a) \Phi / \Phi_{0}$ along 
the $\mp y$ direction, with $\Phi / \Phi_{0}=Ba^{2}e/h$ ($a$ is the lattice 
constant). Disorder is introduced by assigning random fluctuations to the 
orbital energy $\varepsilon_{i}$ , taking $\varepsilon_{i} \leq |W/2|$. In 
the white-noise case, these energies are uncorrelated. In the correlated 
disorder model, a gaussian correlation 
$\varepsilon_{i} = \frac{1}{\pi \lambda^{2}} \sum_{j} \varepsilon_{j} 
e^{-|\mathbf{R}_{i}-\mathbf{R}_{j}|^{2}/\lambda^2}$ [12] with correlation 
length $\lambda$, is assumed.
 The tight-binding parameters are chosen in order to emulate the
effective mass of an electron in the bottom of the GaAs conduction band: 
$m^*=\hbar^2/(2|V|a^2) = 0.067m_{e}$. We focus the analysis on the 
lowest Landau levels for small magnetic fluxes $\Phi/\Phi_0 \!\leq\! 1/20$, 
a range where the lattice effects on the electronic spectra (Landau 
levels) are negligible. We consider squares of 40x40 sites with periodic 
boundary conditions. 

The localization of the states is evaluated by means of the Participation 
Ratio (PR) [14]: 

\begin{equation}
PR=1/(N \sum_{i=1}^{N}|a_{i}|^{4})
\end{equation} 

where $N$ is number of lattice sites and $a_{i}$ is the amplitude of the 
normalized wavefunction on site $i$. The signature of an extended state 
in the $PR$, within each Landau band, will be a sharp peak for sufficient 
large systems. Both, the DOS and PR shown here are averages of 100 disorder 
configurations.

\section{Results and Discussion}

In Fig.1 we show the calculated PR and the DOS for the lowest two broadened 
Landau bands, for a white-noise disorder amplitude of $W/V=2.8$ and magnetic 
flux $\Phi / \Phi_{0}=0.05$. The floating up of the extended state (PR peak 
position) relative to the center of the $1^{st.}$ Landau band, indicated  
by $\delta E$, is evident. 

The extended state actually goes down in energy with increasing disorder, 
due to the overall repulsion of states induced by the potential fluctuation.
The important feature, however, is that this repulsion is more pronounced 
for localized states (as proposed by Haldane and Yang [4]), resulting in a 
relative levitation in energy of the extended states. Fig.2 shows the 
equivalent results to Fig.1 for a correlated disorder of amplitude 
$W/V=4.8$ and a correlation length $\lambda=2a$. Now the levitation of the 
extended states  is not only relative, but absolute in energy. 

\begin{figure}
\epsfxsize=3.2in \epsfbox{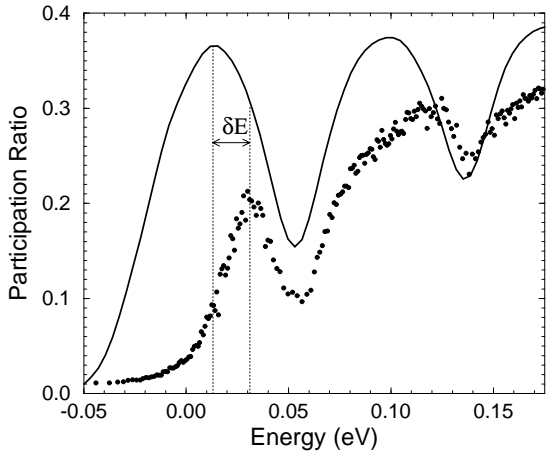}
\caption{DOS (continuous line) and the PR (circles) 
for the lowest two Landau bands for the white-noise disorder model:
$W/V=2.8$ and $\Phi / \Phi_{0}=0.05$. $\delta E$ is the energy shift 
of the extended state.}
\label{1}
\end{figure}

\begin{figure}
\epsfxsize=3.2in \epsfbox{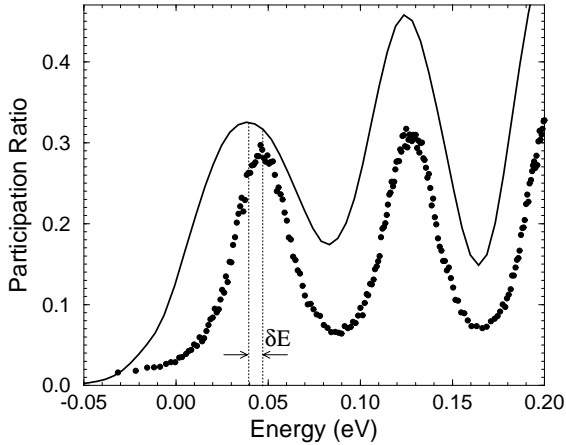}
\caption{Same as Fig.1 for the correlated disorder 
 model with $W/V=4.8$ and $\lambda=2a$.}
\label{2}
\end{figure}
 
From Figs.1 and 2 it becomes clear that changes in the potential landscape 
lead to modifications in both, the DOS and the general features of the 
localization of the states. Concerning the DOS we note that while in the 
white-noise disorder the successive Landau bands have all the same broadening 
$\Gamma$, the correlations make $\Gamma$ dependent on Landau level index: 
the higher the index, the narrower the band [15]. For white-noise potential 
fluctuations, the PR peaks for higher Landau bands become highly asymmetric 
and less resolved than the lowest one, already at relatively small band 
superpositions. On the other hand, a smother potential fluctuation, Fig.2, 
leads to well resolved PR peaks for all Landau bands. Although the shift 
$\delta E$ observed is smaller than that of Fig.1, for approximately the 
same superposition situation, it can be followed until greater superpositions 
and also for higher Landau levels, not only the first one.

\begin{figure}
\epsfxsize=3.2in \epsfbox{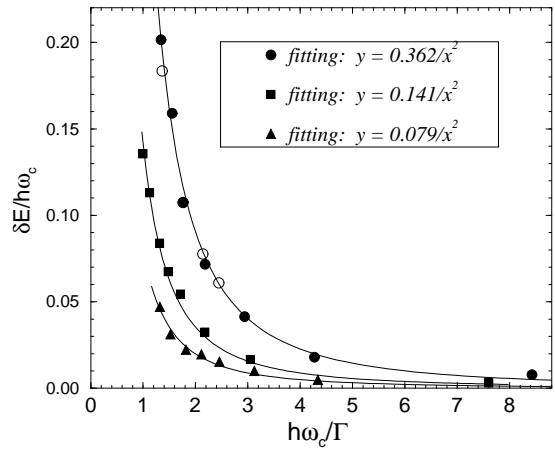}
\caption{Extended states shift {\it vs.} $\hbar\omega_c/\Gamma$. Circles 
 are for the white-noise case. Squares(triangles) are for the first(second)
 extended state for correlated disorder.}
\label{3}
\end{figure}

Referring to the energy shift, $\delta E$, the evolution of the 
critical states as a function of increasing disorder or diminishing 
magnetic field can be followed in Fig.3, which summarizes our extensive 
studies. Here the normalized shift, $\delta E/\hbar\omega_{c}$, are 
plotted as a function of $\hbar\omega_c/\Gamma$, the ratio between the 
energy separation of the DOS peaks and the Landau band width (determined 
at half height). For the white-noise case we follow $\delta E$ by either 
varying the disorder (filled circles) or magnetic flux (open circles). 
Squares and triangles are for the first and second Landau band critical 
states for the gaussian-correlated disordered system, respectively.

The equivalence between diminishing the magnetic field or increasing 
disorder for observing the floating up of the extended states is 
established for the white-noise case. The dependence observed is 
thus a general result valid for any magnetic flux or disorder values 
in the showed $\hbar\omega_c/\Gamma$ range. For the correlated disorder 
it is seen that for the $1^{st}$ Landau level the levitation is less 
pronounced than that of white-noise, for equivalent $\hbar\omega_c/\Gamma$. 
This fact qualitatively agrees with analytical predictions [4, 5] and 
represents the first numerical quantitative comparison of different 
disorder models levitation. Also from Fig.3, a weaker levitation for the
$2^{nd}$ critical state compared to the $1^{st}$ one is verified (notice 
that results shown in Fig.3 already take in account the narrower $\Gamma$
of $2^{nd}$ Landau band, so this is a intrinsic smaller floating up process). This 
second fact contradicts the perturbative approaches [4,5] and initial 
conjectures for levitation [1, 2] that expects more pronounced shifts for 
higher delocalized levels. Finally, it is worth noting that for all cases 
$\delta E/\hbar\omega_{c} \propto (\hbar\omega_c/\Gamma)^{-2}$.

\begin{figure}
\epsfxsize=3.2in \epsfbox{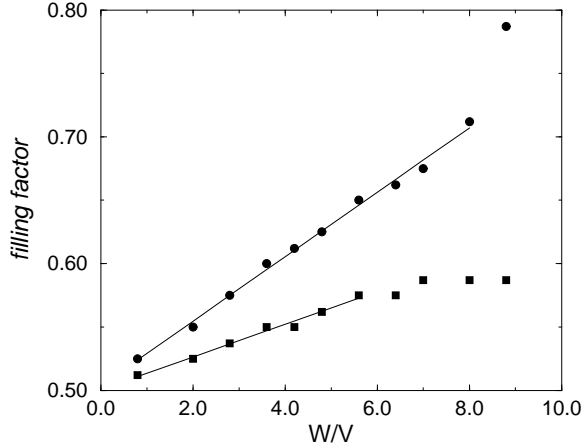}
\caption{Levitation of the filling factor {\it vs.} disorder 
strength for the correlated disorder case. Circles(squares) are for the 
first(second) extended state.}
\label{4}
\end{figure}

In Fig.4 we see the floating up of the filling factors, $\nu$, corresponding 
to the two lowest extended states as a function of disorder strength for the 
correlated disorder case. The values for the $2^{nd}$ extended state have 
been rigidly shifted down by $\nu=1$, in order to better compare them with 
the results for the $1^{st}$ one. Again we can see the less pronounced levitation 
of the $2^{nd}$ extended levels. The striking result is that the floating up is 
linear for a wide range of $W/V$, occurring already before the Landau bands 
superposition, which only starts for $W/V\approx3.6$.  This behavior is 
analogous for the white-noise case (not shown here). The relevance of this 
observation is that it unambiguously contradicts the ``apparent levitation" 
concept, which credits the floating up of the filling factor for the lowest 
extended state exclusively to  Landau band superpositions [6, 16].
Superpositions nevertheless affect the filling factor and indeed
deviations from linear floating up, observed in Fig.4 for strong 
disorder, are due to Landau band superpositions. The levitation for the lowest 
state is enhanced, while the second one shows a saturation, due to the 
broadening dependence on the Landau level index.

\section{Final remarks}

In conclusion, we could quantitatively identify  the levitation of extended 
states in energy, relative to the Landau band centers, for both, white-noise 
and correlated disorder, either as a function of decreasing magnetic field 
or increasing disorder strength. We believe that the present results settle 
the controversy on the levitation in a lattice model: bona fide levitation 
of delocalized states occurs and a correct interpretation 
of the floating-up picture has to focus on the evolution of extended states 
positions relative to the DOS. The well established behavior of the 
extended states for $\hbar\omega_{c}/\Gamma \gtrsim 1$ suggests an intermediate 
situation between the Global Phase Diagram [17] (based on Laughlin conjecture 
[1]) and a recent proposal [8] (non-float-up phase diagram). Extending the 
behavior of a less pronounced levitation of higher extended states to  
$\hbar\omega_{c}/\Gamma < 1$, the present results would also support the 
possibility of direct transitions from $\nu > 1$ states to the insulator phase.
However, the size of the system has to be continuously increased for decreasing 
magnetic field or increasing disorder in order to follow the extended states for
$\hbar\omega_c/\Gamma<1$ [3]. Therefore, the above conclusions can not be 
extended to arbitrarily low magnetic fields, and the implications of these 
results on the Quantum Hall Phase Diagram have to be taken carefully. Perhaps 
more important than the size effects, are the unavoidable lattice effects 
(annihilation of current carrying states with opposite Chern numbers)
[10]. These lattice effects, however, are overcome by the correlation in the
disorder [13] which has the property of widening the ``continuum limit window" of 
lattice models. Finally, the analysis of $\delta E/\hbar\omega_{c}$ $vs.$ 
$\hbar\omega_c/\Gamma$  permits a direct comparison between disorder models or 
even with experimental results including density of states measurements [18].

\section{Acknowledgments}

The authors are grateful to E. R. Mucciolo for several discussions and 
suggestions. Financial support from FAPESP and CNPq are acknowledged.

\end{document}